\documentclass[aps,prl,twocolumn,showpacs,floatfix,superscriptaddress]{revtex4}

\usepackage{graphicx}
\usepackage{dcolumn}
\usepackage{bm}
\usepackage[usenames]{color}    
\usepackage{amsmath}
\usepackage{amsfonts}
\usepackage{amssymb}
\usepackage{graphicx}
\usepackage{bm}
\usepackage{color}

\newcommand{\be}{\begin{equation}}
\newcommand{\ee}{\end{equation}}
\newcommand{\bea}{\vspace{0.25cm}\begin{eqnarray}}
\newcommand{\eea}{\end{eqnarray}}

\begin{document}

\title{Experiment Investigating the Connection between Weak Values and Contextuality}

\author{F.~Piacentini}
\affiliation{INRIM, Strada delle Cacce 91, I-10135 Torino, Italy}

\author{A.~Avella}
\affiliation{INRIM, Strada delle Cacce 91, I-10135 Torino, Italy}


\author{M.~P.~Levi}
\affiliation{INRIM, Strada delle Cacce 91, I-10135 Torino, Italy}

\author{R.~Lussana}

\author{F.~Villa}

\author{A.~Tosi}

\author{F.~Zappa}
\affiliation{Politecnico di Milano, Dipartimento di Elettronica, Informazione e Bioingegneria, Piazza Leonardo da Vinci 32, 20133 Milano, Italy}

\author{M.~Gramegna}

\author{G.~Brida}

\author{I.~P.~Degiovanni}
\affiliation{INRIM, Strada delle Cacce 91, I-10135 Torino, Italy}

\author{M.~Genovese}
\affiliation{INRIM, Strada delle Cacce 91, I-10135 Torino, Italy}
\affiliation{INFN, Via P. Giuria 1, I-10125 Torino, Italy}

\begin{abstract}
Weak value measurements have recently given rise to a large interest for both the possibility of measurement amplification and the chance of further quantum mechanics foundations investigation.
In particular, a question emerged about weak values being proof of the incompatibility between Quantum Mechanics and Non-Contextual Hidden Variables Theories (NCHVT).
A test to provide a conclusive answer to this question was given in [M. Pusey, Phys. Rev. Lett. \textbf{113}, 200401 (2014)], where a theorem was derived showing the NCHVT incompatibility with the observation of anomalous weak values under specific conditions.
In this paper we realize this proposal, clearly pointing out the connection between weak values and the contextual nature of Quantum Mechanics.
\end{abstract}


\maketitle

In 1988 Aharonov, Albert and Vaidman introduced \cite{2} weak value measurements \cite{1,press14,tam13,eli14}, firstly realized in \cite{3,3a,3b,gog10,dre11,gro13,spo15,whi16}, that represent a new paradigm of quantum measurement where so little information is extracted from a single measurement that the state does not collapse.\\
Weak values, i.e. weak measurements of an operator performed on an ensemble of pre- and post-selected states, present non-classical properties, assuming anomalous values (i.e. values outside the eigenvalue range of the observable).
In the recent years they have been subject of a large interest both for the possibility of amplifying the measurement of small parameters \cite{3b,6,7,lun11} and for their non-classical properties, allowing the investigation of fundamental aspects of quantum mechanics \cite{press14,tam13,piac15}.\\
In particular, a question emerged about anomalous weak values constituting a proof of the incompatibility of quantum theory with Non-Contextual Hidden Variables Theories (NCHVT) \cite{spek1,spek2,gen05}, i.e. theories assuming that a predetermined result of a particular measurement does not depend on which other observables are simultaneously measured \footnote{NCHVT are a subset of Local realistic Hidden Variables Theories, assuming that a predetermined result of a particular measurement does not depend on which other observables are simultaneously measured in a causally non-connected region, see M. Genovese, Phys. Rep. 413, 319 (2005)}.
The possibility of testing this connection was recently unequivocally demonstrated in \cite{pusey}, showing that the mere observation of anomalous weak values is largely insufficient for this purpose, while non-contextuality is incompatible with the observation of anomalous weak values under specific experimental conditions.
This result is of deep importance for understanding the role of contextuality in quantum mechanics, also in view of possible applications to quantum technologies.\\
To properly define the connection between non-contextuality and weak values, in ref. \cite{pusey} was presented and proved the following theorem, here in a form avoiding any (hidden) reference to Quantum Mechanics \cite{pusey2}:\\\\

\textbf{Theorem 1.} \textit{Let us suppose to have a preparation procedure $\mathcal{P}_{\psi_i}$, a sharp measurement procedure $\mathcal{M}_{\psi_f}$ with outcomes ``PASS'' and ``FAIL'', and a non-destructive measurement procedure $\mathcal{M}_W$ with outcomes $x\in\mathbb{R}$, such that:
\begin{enumerate}
  \item The pre- and post-selected states $|\psi_i\rangle$ and $|\psi_f\rangle$ are non-orthogonal, i.e.:
      \begin{equation}\label{cond1nc}
        p_{\psi_f} := \mathbb{P}\left(\mathrm{PASS}|\mathcal{P}_{\psi_i},\mathcal{M}_{\psi_f}\right)>0\;;
      \end{equation}
  \item Ignoring the post-measurement state, $\mathcal{M}_W$ is equivalent to a two-outcome measurement with unbiased noise, i.e.:
      \begin{equation}\label{cond2nc}
        \mathbb{P}\left(x|\mathcal{P},\mathcal{M}_W\right) = p_n(x-g)\mathbb{P}\left(1|\mathcal{P},\mathcal{M}_\Pi\right) +$$$$+ p_n(x)\mathbb{P}\left(0|\mathcal{P},\mathcal{M}_\Pi\right)\;\;\;\forall\mathcal{P}
      \end{equation}
      for some sharp measurement procedure $\mathcal{M}_\Pi$ with outcomes ``0'' and ``1'', and probability distribution $F(x)$ with median $x=0$;
  \item We can define a ``probability of disturbance'' $p_d$ such that, ignoring the outcome of $\mathcal{M}_W$, it affects the post-selection in the same way as mixing it with another measurement:
      \begin{equation}\label{cond3nc}
        \mathbb{P}\left(\mathrm{PASS}|\mathcal{P},\mathcal{M}_W,\mathcal{M}_{\psi_f}\right) = (1-p_d)\mathbb{P}\left(\mathrm{PASS}|\mathcal{P},\mathcal{M}_{\psi_f}\right) +$$$$+ p_d\mathbb{P}\left(PASS|\mathcal{P},\mathcal{M}_d\right)\;\;\;\forall\mathcal{P}
      \end{equation}
      for some measurement procedure $\mathcal{M}_d$ with outcomes ``PASS'' and ``FAIL'';
  \item The values of $x$ under the pre- and post-selection have a negative bias that ``outweighs'' $p_d$, i.e. for the quantity $p_-:=\left(p_{\psi_f}\right)^{-1}\int_{-\infty}^0 \mathbb{P}\left(x,\mathrm{PASS}|\mathcal{P}_{\psi_i},\mathcal{M}_W,\mathcal{M}_{\psi_f}\right)\mathrm{d}x$ holds the inequality:
      \begin{equation}\label{cond4nc}
        \mathcal{I}= p_- - \frac12 - \frac{p_d}{p_{\psi_f}} > 0.
      \end{equation}
\end{enumerate}
Then there is no measurement non-contextual ontological model for the preparation $\mathcal{P}_{\psi_i}$, measurement $\mathcal{M}_W$, and
post-selection on ``PASS'' of $\mathcal{M}_{\psi_f}$ satisfying outcome determinism for sharp measurements.\\\\}
\begin{figure}[htbp]
\begin{center}
\includegraphics[width=0.99\columnwidth]{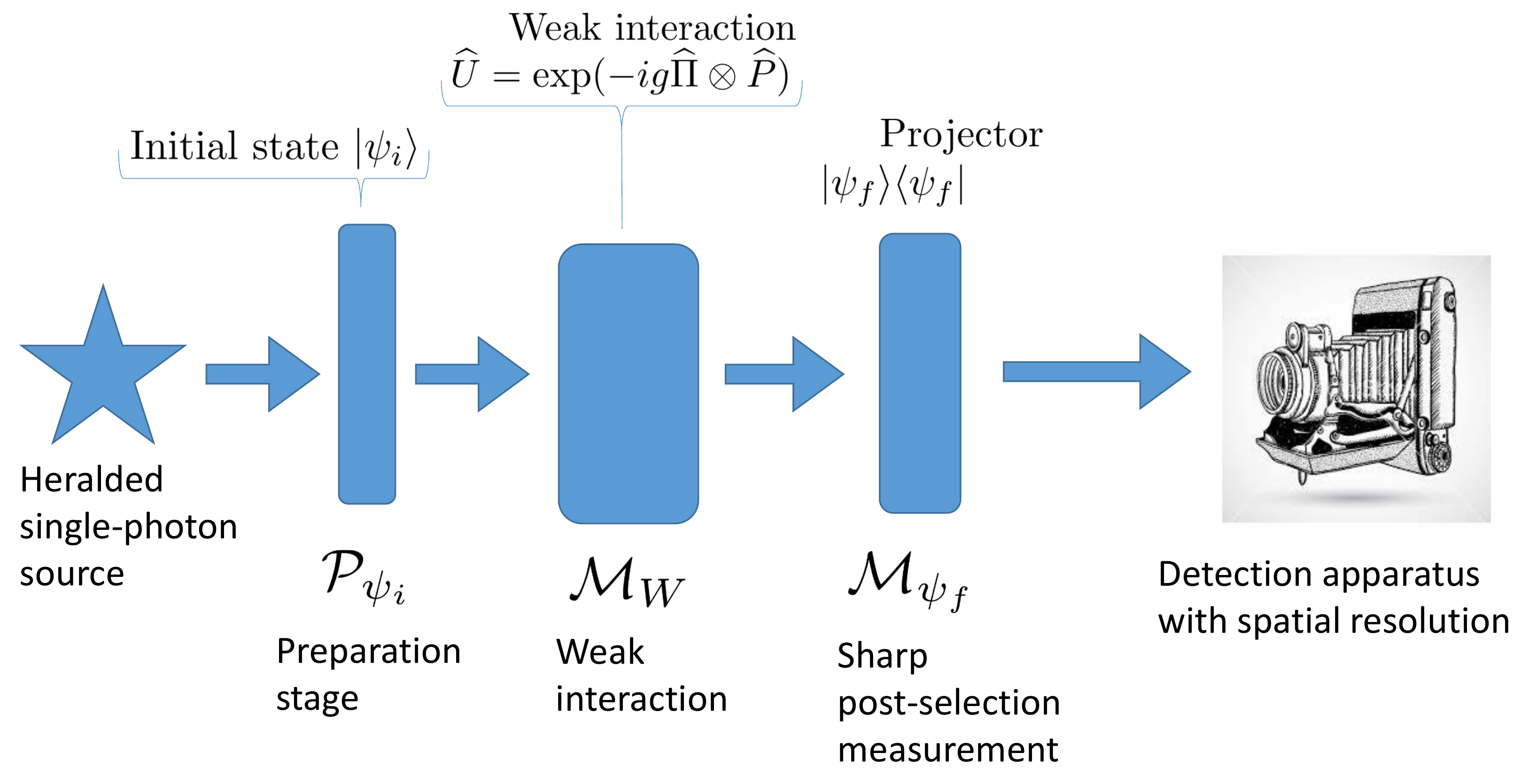}
\caption{Scheme of a ``gedanken'' experiment for the non-contextuality test of theorem 1. The single photons, prepared in the initial state $|\psi_i\rangle$, undergo a weak interaction and a sharp post-selection measurement before being addressed to a detector with spatial resolution. For each part of the scheme, both the non-contextual (below) and quantum mechanical (above) description of its effect are reported.}
\label{NCvsQM}
\end{center}
\end{figure}
Here we present the very first experimental test of this theorem, performed by exploiting polarisation weak measurements on heralded single photons \cite{alan,NHSPS}.\\
In the framework of Quantum Mechanics, the preparation procedure $\mathcal{P}$ corresponds to the pre-selection of the polarisation state $|\psi\rangle=\cos\theta|H\rangle+e^{i\beta}\sin\theta|V\rangle$ of our single photons, while the post-selection process $\mathcal{M}_{\psi_f}$ is represented by the projector $|\psi_f\rangle\langle\psi_f|$, that yields for the probability $p_{\psi_f}$ the equivalence $p_{\psi_f}:=\mathbb{P}\left(\mathrm{PASS}|\mathcal{P},\mathcal{M}_{\psi_f}\right)=\left|\langle\psi_f|\psi\rangle\right|^2$.
The non-destructive measurement procedure $\mathcal{M}_W$, instead, is implemented as a weak interaction induced by the unitary evolution $\widehat{U}=\exp (- i g \widehat{\Pi} \otimes \widehat{P})$, being $g$ the von Neumann coupling constant between the observable $\widehat{\Pi}$ and a pointer observable $\widehat{P}$ (see Fig.\ref{NCvsQM}).\\
In our experiment, a single photon state is prepared in the initial state $|\phi\rangle\rangle=|\psi\rangle\otimes|f_x\rangle$, with $|f_x\rangle = \int \mathrm{d} x F (x) |x \rangle$, where $|F (x)|^2=p_n(x)$ is the probability density function of detecting the photon in the position $x$ of the transverse spatial plane.
The shape of $p_n(x)$ is Gaussian with good approximation, since the single photon guided in a single-mode optical fiber is collimated with a telescopic optical system (see Fig.\ref{setup}), and by experimental evidence we can assume the (unperturbed) $p_n(x)$ to be centered around zero with width $\sigma$.\\
The single photon undergoes a weak interaction realized as a spatial walk-off induced in a birefringent crystal, described by the unitary transformation $\widehat{U}$.
The probability of finding the single photon in the position $x_0$ of the transverse plane (see Eq.(\ref{cond2nc})) can be evaluated as:
\begin{equation}\label{Px0QM}
  \mathbb{P}\left(x_0|\mathcal{P},\mathcal{M}_W\right) = \mathrm{tr}\left[M_{x_0}|\psi\rangle\langle\psi|M^\dagger_{x_0}\right]
\end{equation}
where $M_{x_0}|\psi\rangle=\langle x_0| \widehat{U} |\phi\rangle\rangle$. The quantities $\mathbb{P}\left(1|\mathcal{P},\mathcal{M}_\Pi\right)$ and $\mathbb{P}\left(0|\mathcal{P},\mathcal{M}_\Pi\right)$ in Eq.(\ref{cond2nc}) correspond respectively to the probability that the single photon undergoes or not the weak interaction in the crystal, i.e. $\mathbb{P}\left(1|\mathcal{P},\mathcal{M}_\Pi\right) = \langle\psi|\widehat{\Pi}|\psi\rangle$ and $\mathbb{P}\left(0|\mathcal{P},\mathcal{M}_\Pi\right)= 1-\mathbb{P}\left(1|\mathcal{P},\mathcal{M}_\Pi\right)= \langle\psi|\widehat{\tilde{\Pi}}|\psi\rangle$ (being $\widehat{\tilde{\Pi}}=I-\widehat{\Pi}$).\\
The quantity $\mathbb{P}\left(\mathrm{PASS}|\mathcal{P},\mathcal{M}_d\right)$ in Eq.(\ref{cond3nc}) represents an unknown measurement process, but what we need to demonstrate is just that its contribution is negligible, because of the non-destructive nature of the measurement $\mathcal{M}_W$ (since we exploited the weak measurement paradigm). The parameter $p_d$, quantifying such contribution (i.e. the disturbance that $\mathcal{M}_W$ causes to the subsequent sharp measurement $\mathcal{M}_{\psi_f}$) can be evaluated as the amount of decoherence induced on the single photon by the weak interaction $\widehat{U}$, $p_d=1-e^{-\frac{g^2}{4\sigma^2}}$.\\
Our experimental setup (Fig.\ref{setup}) consists of a 796~nm mode-locked Ti:Sapphire laser (repetition rate: 76 MHz), whose second harmonic emission pumps a $10\times10\times5$ mm LiIO$_3$ nonlinear crystal, producing Type-I Parametric Down-Conversion (PDC).
\begin{figure}[tbp]
\begin{center}
\includegraphics[width=0.99\columnwidth]{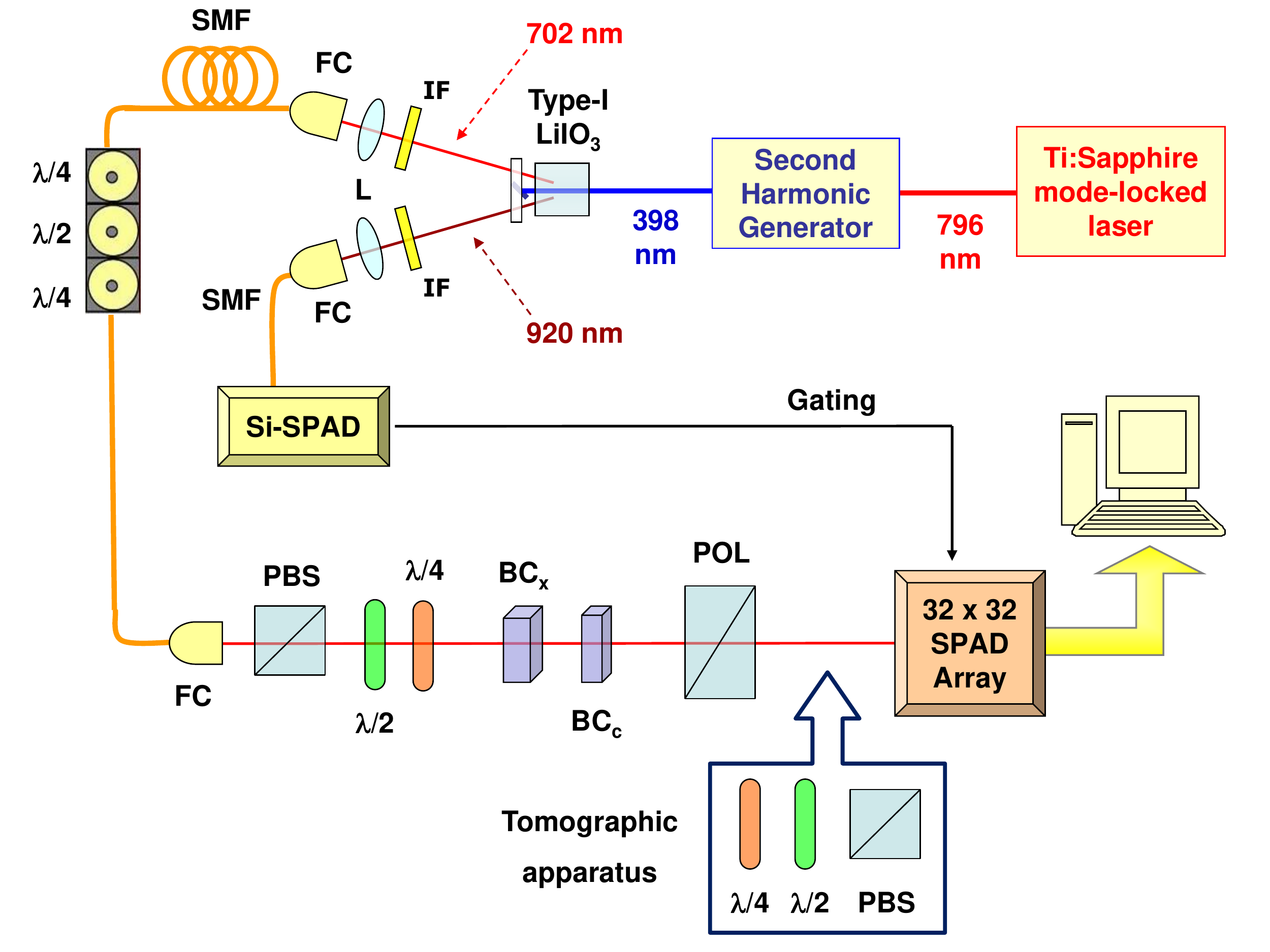}
\caption{(a) Experimental setup. After the weak interaction with a birefringent crystal, the heralded single photon is projected onto the post-selected state by a Glan linear polarizer, and then addressed to the space-resolving detector.
For some consistency checks, a tomographic apparatus can be inserted at some point between the polarizer and the detector.
SHG: Second Harmonic Generator; PBS: Polarizing Beam Splitter; BC: Birefringent Crystal; POL: Glan polarizer).}
\label{setup}
\end{center}
\end{figure}
The idler photon ($\lambda_i=920$ nm) is coupled to a single-mode fiber (SMF) and then addressed to a Silicon Single-Photon Avalanche Diode (SPAD), heralding the presence of the correlated signal photon ($\lambda_s=702$ nm) that, after being SMF-coupled, is sent to a launcher and then to the free-space optical path where the weak values evaluation is performed.\\
We have estimated the quality of our single-photon emission, obtaining a $g^{(2)}(0)$ value (or more properly a parameter $\alpha$ value \cite{h,grangier,bri1})
of $0.13\pm0.01$ without any background/dark-count subtraction.\\
After the launcher, the heralded single photon state is collimated by a telescopic system, and then prepared (pre-selected) in the chosen state $|\psi_{i}\rangle$ by means of a calcite polarizer followed by a quarter-wave plate and a half-wave plate.
The weak measurement is carried out by a 1 mm long birefringent crystal (BC$_x$), whose extraordinary ($e$) optical axis lies in the $X$-$Z$ plane, with an angle of $\pi/4$ with respect to the $Z$ direction.
Due to the spatial walk-off experienced by the vertically-polarized photons, horizontal- and vertical-polarization paths get slightly separated along the $X$ direction, inducing in the initial state $|\psi_{i}\rangle$ a small decoherence (below $1\%$) that keeps it substantially unaffected.
Subsequently, the birefringent crystal BC$_c$ performs a phase compensation tuned in order to nullify the temporal walk-off generated in BC$_x$.
From the parameters $g$ and $\sigma$ of our system, we estimated $p_d=0.0019\pm0.0002$.\\
After the weak measurement is performed, the photon meets a Glan polarizer projecting it onto the post-selected state $|\psi_{f}\rangle$.
Then, the photon goes to the detection device, a two-dimensional array made of $32\times32$ ``smart pixels'', fabricated in a cost-effective 0.35 $\mu$m standard CMOS technology.
Each pixel hosts a 30 $\mu$m diameter silicon SPAD detector with $15\%$ Photon Detection Efficiency (PDE) at 702 nm (peak PDE is $55\%$ at 420 nm), and its front-end electronics for sensing and quenching the avalanche and counting the number of detected photons \cite{villa2014}.
The SPADs are gated with 6 ns integration windows, triggered by the SPAD detector of the heralding arm; spurious detections within such integration windows are minimized thanks to the array's excellent Dark Counting Rate (DCR) performance (120 cps at room temperature, with just $3\%$ hot pixels).\\
A removable polarization tomographic apparatus \cite{kwiat,genotom} is inserted between the Glan polarizer and the detector only when needed, i.e. to verify the fulfillment of the condition in Eq. (\ref{cond3nc}).\\
In Fig.\ref{viol} is reported the plot of the quantity $\mathcal{I}$ of Eq. (\ref{cond4nc}) with respect to the angle $\theta$ of the linearly polarized post-selection state $|\psi_f\rangle=\cos(\theta)|H\rangle+\sin(\theta)|V\rangle$, with $|\psi_i\rangle=\frac{1}{\sqrt{2}}\left(|H\rangle-|V\rangle\right)$. Experimentally, by choosing $\theta=0.18\pi$ we obtained the value $\mathcal{I}^{(exp)}=0.063\pm0.011$, in excellent agreement with the quantum-mechanical predictions and 5.7 standard deviations distant from the non-contextual bound.\\
\begin{figure}[tbp]
\begin{center}
\includegraphics[width=0.95\columnwidth]{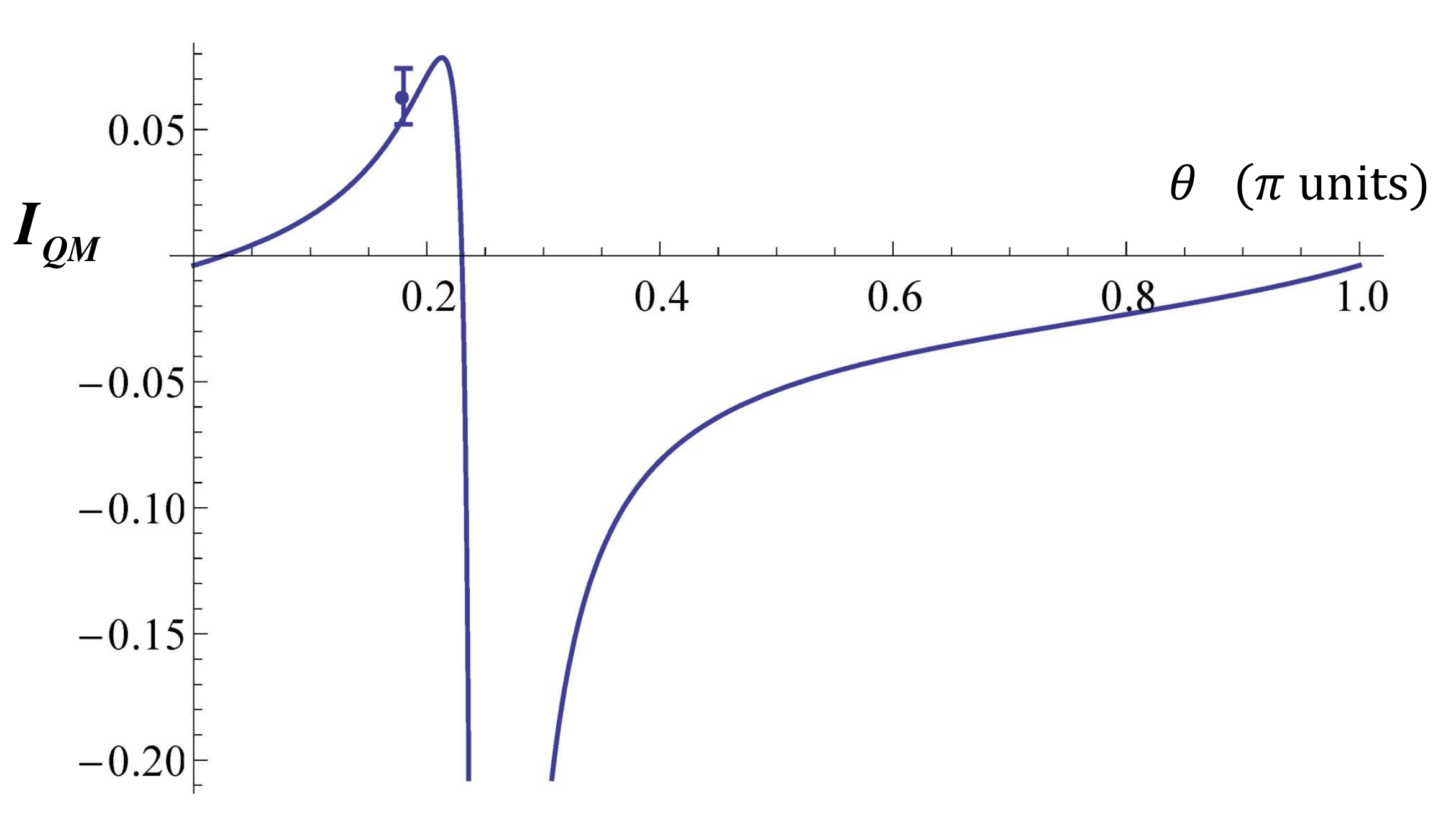}
\caption{$\mathcal{I}$ plot, with respect to the post-selection angle $\theta$. For $\theta=0.18\pi$, we obtained $\mathcal{I}^{(exp)}=0.063\pm0.011$, certifying a violation of the non-contextual bound of 5.7 standard deviations.}
\label{viol}
\end{center}
\end{figure}
In order to demonstrate the validity of Eq.(\ref{cond2nc}), we removed the polarizer realising $\mathcal{M}_{\psi_f}$, so that we could estimate the probability $Q(x)$ that a single photon prepared in any arbitrary polarisation state $|\psi\rangle$ is detected at the position $x$ after the weak interaction, a faithful estimation of $\mathbb{P}\left(x|\mathcal{P},\mathcal{M}_W\right)$.
This task was accomplished by sending the (tomographically complete) set of four different input states $\big\{|H\rangle$, $|V\rangle$, $|+\rangle=\frac{1}{\sqrt{2}}\left(|H\rangle+|V\rangle\right)$, $|R\rangle=\frac{1}{\sqrt{2}}\left(|H\rangle-i|V\rangle\right)\big\}$, and measuring $Q(x)$ in absence of the polarizer performing the state post-selection.
Then, we compared the measured $Q(x)$ with the expected one obtained from the right side of Eq.(\ref{cond2nc}); the function $p_n(x)$ is reconstructed by fitting the spatial profile in absence of the weak interaction ($\mathbb{P}\left(1|\mathcal{P},\widehat{\Pi}\right)=0$), and the value of $g$ is estimated maximizing the interaction ($\mathbb{P}\left(0|\mathcal{P},\widehat{\Pi}\right)=0$).\\
The validity of our approach is shown by the fidelity between the measured $Q(x)$ and the expected one $Q^{(e)}(x)$, evaluated by sampling more than 230 points in the region where $Q(x)$ is significantly non-zero, obtaining 0.997, 0.991, 0.994, 0.996 for the four input states $|H \rangle$, $|V\rangle$, $|+ \rangle$ and $| R \rangle$, respectively.
To confirm the quality of our reconstruction, we also performed a pixel-by-pixel proximity test of the two probability distributions for the pixels where the $Q(x)$ is significantly non-zero.
We define the proximity between the two distributions as:
\begin{equation} \label{Prox}
PROX_{\psi}(x)= \left[\frac{2 Q(x) Q^{(e)}(x)}{ (Q(x))^2 + (Q^{(e)}(x))^2}\right]^{\frac12} .
\end{equation}
As shown in Fig.\ref{fig4}, for all the input states the proximity between the two distributions is larger than 0.99 for almost every point, demonstrating that our experimental setup provides a faithful realization of the condition in Eq.(\ref{cond2nc}).\\
\begin{figure}[tbp]
\begin{center}
\includegraphics[width=0.99\columnwidth]{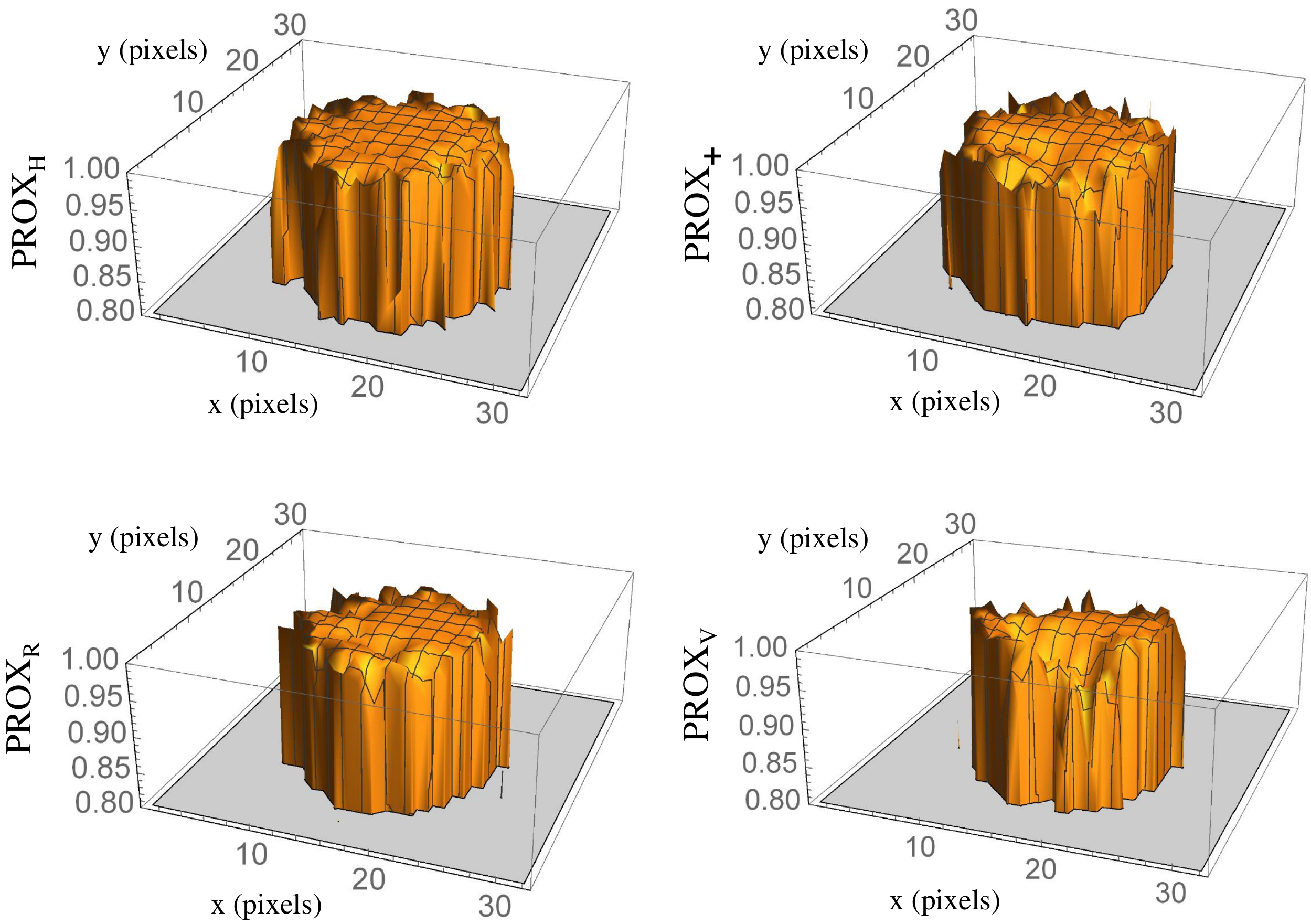}
\caption{$PROX_{\psi}(x)$ evaluated for the  input states $|\psi\rangle = |H \rangle $, $|+ \rangle$, $|R \rangle$ and $|V\rangle$. For all of them, the proximity is above 0.99 for almost all the pixels, clear sign that the condition of Eq.(\ref{cond2nc}) is satisfied.}
\label{fig4}
\end{center}
\end{figure}
Finally, to prove that the condition of Eq.(\ref{cond3nc}) is fulfilled, we used the following method, based on the comparison between experimental probabilities collected in different conditions, in order to get rid of any possible bias due to quantum mechanical assumptions.
First, we prepared a (tomographically complete) set of states and registered the detection probabilities $P$ and $\tilde{P}$, obtained with the Glan polarizer projecting the single photon states onto $|\psi_f\rangle$ and its orthogonal $|\tilde{\psi}_f\rangle$.
For each input state $|\psi\rangle$, these probabilities are given by the photon counts divided by the trigger counts $N_T$ of the heralded single photon source ($P=\frac{N^{\psi_f}}{N_T}$, $\tilde{P}=\frac{N^{\tilde{\psi}_f}}{N_T}$).
Second, we switched the position of the preparation stage and the birefringent crystals, in order to nullify the weak interaction without altering the optical losses in the system, and performed the same set of acquisitions.
To get rid of any bias, proper dark counts and background noise subtraction is performed.\\
For each input state, these two acquisitions correspond respectively to the evaluation of the quantities $\mathbb{P}\left(\mathrm{PASS}|\mathcal{P},\mathcal{M}_W,\mathcal{M}_{\psi_f}\right)$ and $\mathbb{P}\left(\mathrm{PASS}|\mathcal{P},\mathcal{M}_{\psi_f}\right)$ reported in Eq.(\ref{cond3nc}).
Concerning the third one, connected to the unknown measurement procedure $\mathcal{M}_d$, one can notice that by definition $\mathbb{P}\left(\mathrm{PASS}|\mathcal{P},\mathcal{M}_{d}\right)\in[0,1]$, and thus one can write
\begin{equation}\label{ineq_pd}
  (1-p_d)\mathbb{P}\left(\mathrm{PASS}|\mathcal{P},\mathcal{M}_{\psi_f}\right) \leq \mathbb{P}\left(\mathrm{PASS}|\mathcal{P},\mathcal{M}_W,\mathcal{M}_{\psi_f}\right) \leq $$$$ \leq (1-p_d)\mathbb{P}\left(\mathrm{PASS}|\mathcal{P},\mathcal{M}_{\psi_f}\right) + p_d,
\end{equation}
giving an upper and lower bound to the parameter $p_d$.
The collected data allowed us to obtain $(0.000021\pm0.000014) \leq p_d \leq (0.086\pm0.050)$; the $p_d$ value derived by the system parameters fits perfectly in this range.\\
As a further consistency check, we tested the output state after the sharp measurement $\mathcal{M}_{\psi_f}$ (realized by the Glan polarizer) by inserting the tomographic apparatus in the setup (see Fig.\ref{setup}), implicitly accepting some quantum mechanical assumptions.
Such apparatus was exploited to perform two different experiments.\\
In the first one, we used it to project the state after $\mathcal{M}_{\psi_f}$ onto $\psi_f$ and $\tilde{\psi}_f$.
While we were able to detect a clear signal with the tomographic device realizing the same projection as the Glan polarizer (i.e. onto $\psi_f$), the amount of signal registered with the tomographer projecting onto $\tilde{\psi}_f$ was so small to be completely indistinguishable from the detector noise, as expected when photons undergo two subsequent projections onto orthogonal axes.
This confirms that the sharp measurement process $\mathcal{M}_{\psi_f}$ is performing a projection onto the state $\psi_f$.\\
In the second experiment, instead, we performed the tomographic reconstruction of the state after the post-selection on $|\psi_f \rangle$.
We prepared a tomographically complete set of input states, i.e. $|H \rangle$, $|+ \rangle$, $|L\rangle=\frac{1}{\sqrt{2}}\left(|H\rangle+i|V\rangle\right)$ and $|R\rangle$, and tried to reconstruct via quantum tomography the state after the $\mathcal{M}_{\psi_f}$ measurement process.
From the tomographic reconstructions we obtained states whose fidelities with respect to the chosen $|\psi_f\rangle$ were $\mathcal{F}_H =0.9995 $, $\mathcal{F}_+ = 0.9999$, $\mathcal{F}_L =0.9991$, $\mathcal{F}_R =0.9811$.
These values lead to estimate $p_d=0.0051\pm0.0046$, fitting the range obtained for $p_d$ with the method presented above and in good agreement with the $p_d$ value derived from the system experimental parameters ($p_d=0.0019\pm0.0002$).\\
Since all the conditions of the theorem presented in \cite{pusey} have been verified, we can assess that the results of our experiment clearly violate the non-contextual bound for the quantity $\mathcal{I}$ in Eq.(\ref{cond4nc}), providing a sound demonstration of the connection between weak values and the intrinsic contextual nature of Quantum Mechanics.\\\\

\textbf{Acknowledgments}
This work has been supported by EMPIR-14IND05 ``MIQC2'' (the EMPIR initiative is co-funded by the EU H2020 and the EMPIR Participating States), and by John Templeton Foundation (Grant ID 43467).\\
We are deeply indebted with Matthew Pusey for fruitful discussions and theoretical support.

\end{document}